\documentclass[12pt,english,floatfix,superscriptaddress,aps,prd,preprint,onecolumn,showpacs]{revtex4}
\usepackage{amsmath}
\usepackage{amssymb}
\usepackage{amsbsy}
\usepackage{amsfonts}
\usepackage{amsopn}
\usepackage{amstext}
\usepackage{graphicx}
\usepackage{amssymb}
\usepackage{amsfonts}
\usepackage{amsmath}
\usepackage{graphicx}
\usepackage[english]{babel}
\usepackage{color}
\usepackage{esint}
\usepackage[dvips]{epsfig}
\usepackage[dvips]{graphicx}
\usepackage{float}
\usepackage{units}
\usepackage{textcomp}

\usepackage{hyperref}
\usepackage{slashed}

\usepackage{graphicx} 
\usepackage{subfigure} 

\DeclareMathOperator{\e}{e}
\DeclareMathOperator{\sech}{sech}

\begin{document}

\title{{Gravity localization on hybrid branes}}

\author{D. F. S. Veras}
\email{franklin@fisica.ufc.br}
\affiliation{Universidade Federal do Cear\'a (UFC), Departamento de F\'isica, Campus do Pici, Fortaleza - CE, C.P. 6030, 60455-760 - Brazil}


\author{W. T. Cruz}
\email{wilamicruz@gmail.com}
\affiliation{Instituto Federal de Educa\c{c}\~{a}o, Ci\^{e}ncia e Tecnologia do Cear\'{a} (IFCE), Campus Juazeiro do Norte - 63040-540 Juazeiro do Norte-Cear\'{a}-Brazil}


\author{R. V. Maluf}
\email{r.v.maluf@fisica.ufc.br}
\affiliation{Universidade Federal do Cear\'a (UFC), Departamento de F\'isica, Campus do Pici, Fortaleza - CE, C.P. 6030, 60455-760 - Brazil}


\author{C. A. S. Almeida}
\email{carlos@fisica.ufc.br}
\affiliation{Universidade Federal do Cear\'a (UFC), Departamento de F\'isica, Campus do Pici, Fortaleza - CE, C.P. 6030, 60455-760 - Brazil}

\date{\today}

\begin{abstract}
This work deals with gravity localization on codimension-1 brane worlds engendered by compacton-like kinks, the so-called \textit{hybrid branes}. In such scenarios, the thin brane behaviour is manifested when the extra dimension is outside the compact domain, where the energy density is non-trivial, instead of asymptotically as in the usual thick brane models. 
 The zero mode is trapped in the brane, as required. The massive modes, although are not localized in the brane, have important phenomenological implications such as corrections to the Newton's law. We study such corrections in the usual thick domain wall and in the hybrid brane scenarios. By means of suitable numerical methods, we attain the mass spectrum for the graviton and the corresponding wavefunctions. The spectra possess the usual linearly increasing behaviour from the Kaluza-Klein theories. Further, we show that the $4D$ gravitational force is slightly increased at short distances. The first eigenstate contributes highly for the correction to the Newton's law. The subsequent normalized solutions have diminishing contributions. Moreover, we find out that the phenomenology of the hybrid brane is not different from the usual thick domain wall. The use of numerical techniques for solving the equations of the massive modes is useful for matching possible phenomenological measurements in the gravitational law as a probe to warped extra dimensions.
\end{abstract}


\maketitle

\section{Introduction}
\label{Sec_Introduction}

\textit{Solitons} are structures found as solutions of certain classes of non-linear differential equations which arise from interactions of non-linear and dispersive effects in the medium \cite{Ablowitz, Rajaraman}. They are present in many physical contexts, such as fiber optics \cite{FiberOptics}, protein and polyethylene chains \cite{Davydov, Polyethylene1, Polyethylene2}, DNA macromolecule \cite{Yakushevich}, plasmas \cite{Plasma}, Josephson junctions \cite{JosephsonJunction} and many others. In field theory, the topological defects (solutions which are stable against decays to trivial solutions) usually appear in models that support spontaneous symmetry breaking. The most known examples are kinks, domain walls, vortex, strings and monopoles \cite{Vilenkin}.

An interesting solitonic solution is the \textit{compacton}, found by Rosenau and Hyman \cite{Rosenau-Hyman-1993} as solutions of a special class of the Korteweg-de Vries (KDV) equation. Such structures have compact support. They differ from trivial solution only in a finite region of space. Compactons are found in a wide variety of physical systems where non-linear dispersion arises naturally. For example, the equation governing the motion of a dense chain is a prototype of compacton supporting systems \cite{Rosenau-1994, Gaeta}. Moreover, a non-linear dynamical model of the DNA macromolecule can also  support topological compactons \cite{Compacton-DNA}. Recently, it was observed the existence of compacton matter waves in Bose-Einstein condensates in deep optical lattices \cite{Bose-Einstein}.

Investigation of the presence of compactons in relativistic scalar field theory was also performed \cite{Dussuel-1998, Bazeia-NewResults, Bazeia-FromKinks, Bazeia-CompactStructures}. The $\phi^{4}$ model with non-linear coupling can exhibit a static compact solution \cite{Dussuel-1998}. Besides, non-linear dispersion gives rise to compact structures in models described by a single real scalar field in two-dimensional space-time \cite{Bazeia-NewResults, Bazeia-FromKinks, Bazeia-CompactStructures}.

The physical interest of compactons lies in that they are solitary waves whose energy is strictly localized. Besides, differently from the ordinary solitonic waves which have infinite tails, two compacton-like structures would only interact with each other at the moment of collision \cite{Rosenau-Hyman-1993}.

In the last decade, the braneworld models have gained a considerable amount of attention in the literature, since they provide a solution for the gauge hierarchy and the cosmological constant problems \cite{ADD, AADD, RS1}. In particular, the Randall-Sundrum (RS) model brought the  idea of an infinity extra dimension through a warped geometry \cite{RS2}. Furthermore, the proposal of branes being generated by topological defects was also introduced, wherein domain walls have been used to represent the brane \cite{Rubakov, Gremm}. In this way, the five-dimensional gravity is coupled to background scalar fields \cite{Gremm, DeWolfe}. Such scalar fields extend the singular thin brane in the RS model to the so-called \textit{thick branes} \cite{Csaki-UniversalAspects}. The inclusion of a brane thickness gives new  possibilities  and  new  richer variety of brane worlds \cite{ThickBrane}.

Recently, Bazeia \textit{et al}  developed a mechanism that smoothly interpolates from kinks to compactons in the context of a relativistic field theory \cite{Bazeia-FromKinks}.  Such mechanism made possible the set up of a braneworld scenario being generated by a compacton-like defect. The resulting thick brane possesses a hybrid profile (and hence termed \textit{hybrid brane}): while the usual thick branes behave as thin branes asymptotically, the hybrid brane behaves as a thin one when the extra dimension is outside the domain where the energy density is non-trivial \cite{Bazeia-FromKinks}. 

An important reason to include compacton-like structures to conceive a thick braneworld scenario, lays in that its thickness can be controlled, unlike the ordinary domain walls models \cite{ThickBrane}. In this work, we study the gravity fluctuations in hybrid branes. We show that only one of the models proposed in the Ref. \cite{Bazeia-FromKinks} conducts to a warped spacetime that presents the hybrid profile more clearly. As required, the massless mode (responsible to reproduce the $4D$ gravitational law) is trapped in the brane. Besides, we attain the Kaluza-Klein mass spectrum and the correspondent massive modes. Although the massive modes are not localized in the brane, they have important phenomenological implication: the correction of the Newton's law at short distances. We show that the first massive eigenstate has the highest contribution. The subsequent normalized solutions have diminishing contributions. Moreover the phenomenology is the same for both hybrid and thick domain wall branes.

\section{The hybrid brane scenario} 
\label{Sec-CompactBrane}

In this section, we will present the formalism for building a braneworld scenario in a five dimensional warped spacetime engendered by a compacton-like defect. Let us start from the lagrangian density for a real scalar field $\phi(x)$ with spontaneous symmetry breaking in dimensionless form:
\begin{equation}
\label{Kink-lagrangeano}
\mathcal{L}_k = \frac{1}{2} \partial_{\mu}\phi\partial^{\mu}\phi - V_k(\phi),
\end{equation}
where
\begin{equation}
\label{Potencial-phi4}
V_k(\phi) = \frac{1}{2}(1 - \phi^2)^2.
\end{equation}
The topological solution connecting the minima $\phi_0 = \pm 1$ is the well-known \textit{kink solution}, given explicitly by $\phi(x) = \tanh x$. Its energy density is $\rho(x) = \sech^4 x$.

\textit{Compacton-like} defects appear in relativistic field theory from the Lagrangian \cite{Bazeia-FromKinks}
\begin{equation}
\label{Compacton-lagrangeano}
\mathcal{L}_c = -\frac{1}{4} \left(\partial_{\mu}\phi\partial^{\mu}\phi\right)^2 - \frac{3}{2}V_k(\phi),
\end{equation}
whose equation of motion is
\begin{equation}
\label{Compacton-Eq.Mov}
\left( \frac{d\phi}{dx}\right)^{2} \frac{d^2 \phi}{dx^2} = - \phi (1 - \phi^2).
\end{equation}
In this model, non-linearity raises from the potential and the generalized kinematics introduces non-linear dispersion \cite{Bazeia-FromKinks}. The topological solution is given by \cite{Bazeia-FromKinks}
\begin{equation}
\label{Compacton-solution}
\phi_c(x) =
\left\{
	\begin{array}{ll}
		    -1  & \hspace{0.5cm} \mbox{if} \hspace{0.3cm}  x < \pi/2 \, , \\
		\sin x  & \hspace{0.5cm} \mbox{if} \hspace{0.3cm} |x| \leq \pi/2 \, , \\
		      1 & \hspace{0.5cm} \mbox{if} \hspace{0.3cm}  x > \pi/2 \, .
	\end{array}
\right.
\end{equation}
This solution is stable and has the energy density \cite{Bazeia-FromKinks}
\begin{equation}
\label{Compacton-energy}
\rho_c(x) =
\left\{
	\begin{array}{ll}
		\cos^4 x  & \hspace{0.5cm} \mbox{if}  \hspace{0.5cm} |x| \leq \pi/2 \, ,\\
		         0 & \hspace{0.5cm} \mbox{if} \hspace{0.5cm} |x| > \pi/2 \, .
	\end{array}
\right.
\end{equation}
Note that the solution $\phi_c$ and its energy density have localized support exhibiting the compacton-like structure of the model.

\begin{figure}[!h] 
       \begin{minipage}[t]{0.48 \linewidth}
           \includegraphics[width=\linewidth]{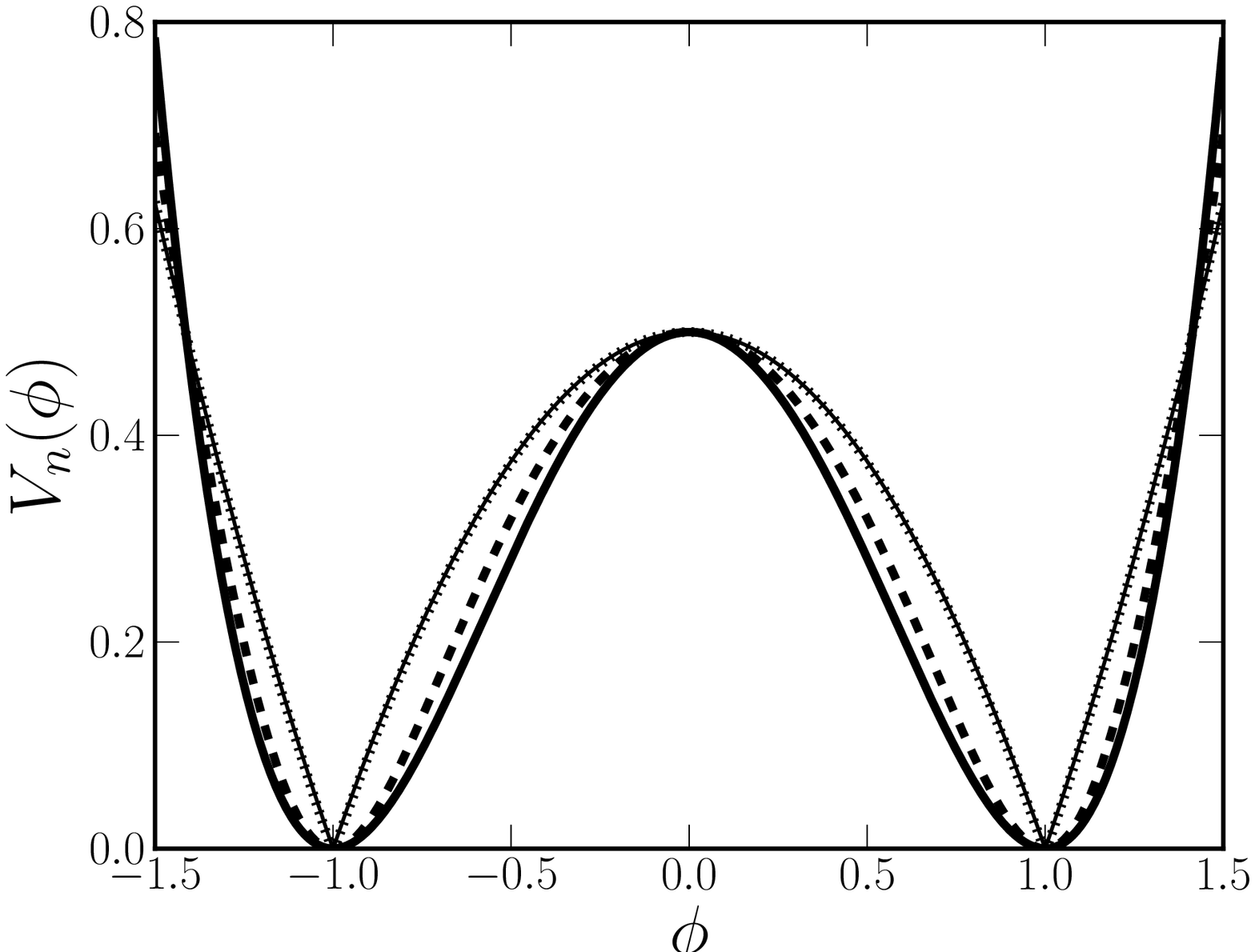}\\
           \caption{Plot of the potential $V_{\alpha}$ for $\alpha = 0.001$ (thick line), $\alpha = 1.0$ (dashed line), $\alpha = 100.0$ (dotted line) and $\alpha = 10000.0$ (thin line). }
          \label{Fig_Valpha}
       \end{minipage}\hfill
       \begin{minipage}[t]{0.48 \linewidth}
           \includegraphics[width=\linewidth]{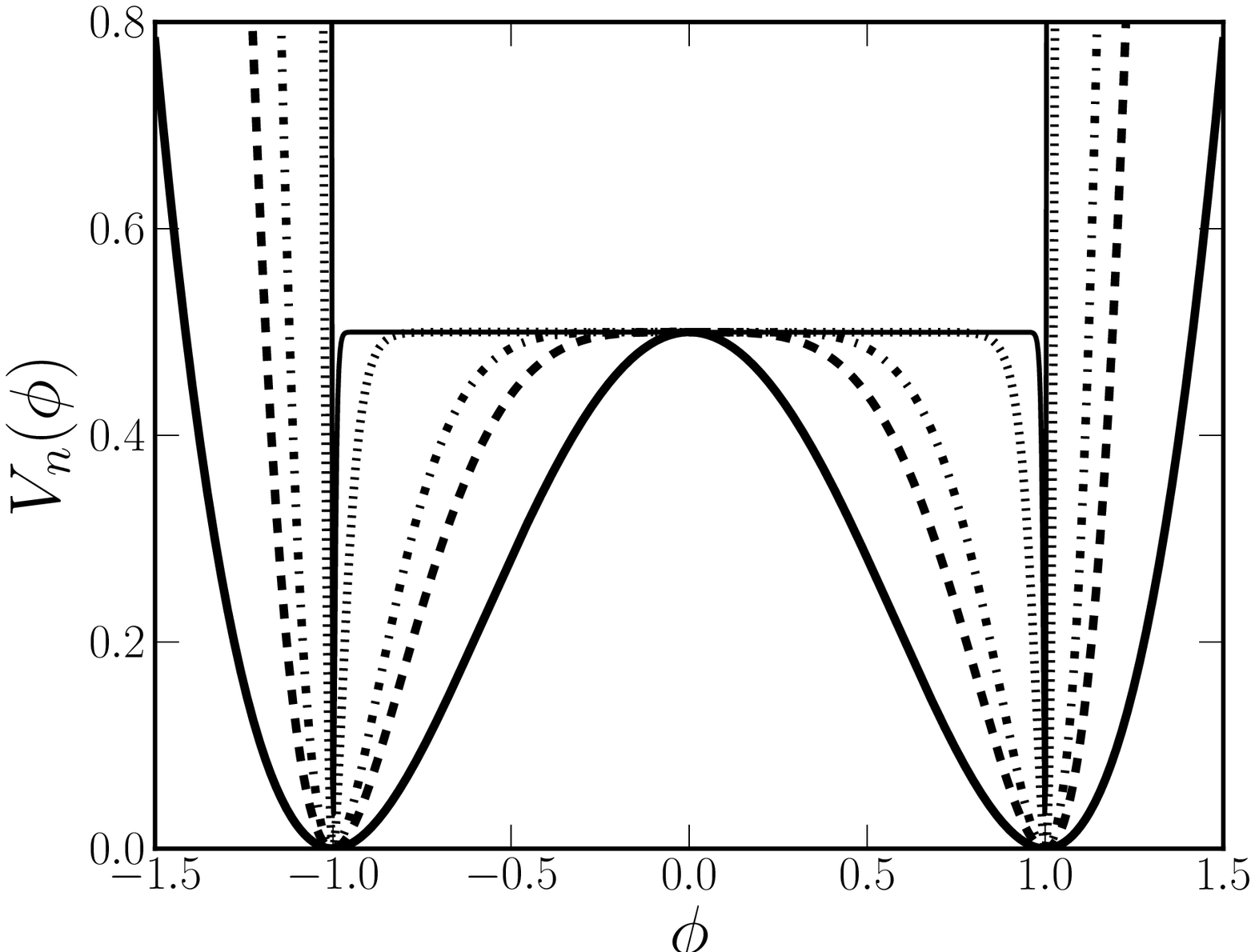}\\
           \caption{Plot of the potential $V_{n}$ for $n = 1$ (thick line), $n = 2$ (dashed line), $n = 3$ (dash-dot line), $n = 15$ (dotted line) and $n = 100$ (thin line).}
           \label{Fig_Vn}
       \end{minipage}
   \end{figure}

Although the two scenarios are very distinct, a method that smoothly transforms kinks into compactons was proposed in Ref. \cite{Bazeia-FromKinks}. In the deformation mechanism \cite{DeformedDefects}, two distinct models with standard kinematics were used \cite{Bazeia-FromKinks}:
\begin{equation}
\label{Lagrangeano}
\mathcal{L}_{\alpha ,n} = \frac{1}{2}\partial_{\mu}\phi\partial^{\mu}\phi - V_{\alpha , n}(\phi),
\end{equation}
where
\begin{equation}
\label{V_alpha}
V_{\alpha}(\phi) = \frac{1}{2\alpha} \left( \sqrt{1 + 4\alpha\left( 1+ \frac{\alpha}{2} \right)V_k(\phi)} - 1 \right)
\end{equation}
and
\begin{equation}
\label{V_n}
V_n(\phi) = \frac{1}{2}\left( 1 - \phi^{2n}\right)^2,
\end{equation}
with $\alpha$ being a non-negative real parameter and $n \geq 1$, an integer one. We plot in Figs. \ref{Fig_Valpha} and \ref{Fig_Vn}, the potentials $V_{\alpha}$ and $V_n$, respectively, for different values of the parameters $\alpha$ and $n$. Note that for $n = 1$ and in the limit $\alpha \rightarrow 0$, the usual $\phi^4$ potential is recovered. The minima and maxima are unaltered. The deformation in $V_{\alpha}$ is very slow if compared with $V_n$. Notice also that, even for $\alpha=100$ or $\alpha=10000$, $V_{\alpha}$ is only slightly modified. Furthermore, note that $V_n$ acquires a compacted shape for larger $n$.


An interesting subject is how to construct a braneworld scenario engendered by a compacton defect. In the Ref. \cite{Bazeia-FromKinks} it was performed for the model $\mathcal{L}_n$ only. We, in turn, will investigate the brane set up mechanism for both models $\mathcal{L}_{\alpha}$ and $\mathcal{L}_{n}$ numerically. 

Consider the scalar field $\phi_{\alpha, n}(y)$ coupled to gravity in an five-dimensional warped spacetime with an extra dimension $y$ of infinite extent. As usual, the metric is  $ds^2 = \e^{2A(y)}\eta_{\mu\nu}dx^{\mu}dx^{\nu} - dy^2$, where $\eta_{\mu\nu}$ is the Minkowski metric with signature $(-,+,+,+)$ and $\e^{2A(y)}$ is the warp factor \cite{ThickBrane}. Thus, the Einstein-Hilbert action reads
\begin{equation}
\label{Einstein-Hilbert}
S_{\alpha,n} = - \int d^5x \, \sqrt{-G}\left( \frac{1}{4}R - \mathcal{L}_{\alpha, n} \right),
\end{equation}
where $R$ is the scalar curvature. To build the braneworld scenario we will made use of the method developed in Ref. \cite{Branalizacao} wherein a superpotential $W(\phi)$ is constructed from the potential $V(\phi)$ in order to reduce the second order equations of motion to first order ones. In this way, the equations of motion read
\begin{equation}
\label{Eq-mov1}
\phi^{\prime\prime} + 4\phi^{\prime}A_{\alpha,n}^{\prime} = \frac{dV_{\alpha,n}}{d\phi},
\end{equation}
\begin{equation}
\label{Eq-mov2}
A_{\alpha,n}^{\prime\prime} = -\frac{2}{3} \left( \phi^{\prime} \right)^2,
\end{equation}
\begin{equation}
\label{Eq-mov3}
\left( A_{\alpha,n}^{\prime} \right)^2 = \frac{1}{6} \phi^{\prime} - \frac{1}{3}V_{\alpha,n}(\phi),
\end{equation}
where the primes denote derivatives with respect to $y$. Defining \cite{Bazeia-FromKinks} 
\begin{equation}
\label{Superpotential}
V_{\alpha,n}(\phi) = \frac{1}{2} \left( \frac{dW_{\alpha,n}}{d\phi}\right)^2,
\end{equation}
the first order differential equations
\begin{equation}
\label{FirstOrder_EqMotion}
\phi_{\alpha, n}^{\prime} = \frac{dW_{\alpha,n}}{d\phi}, \hspace{0.5cm} \text{and} \hspace{0.5cm}  A_{\alpha,n}^{\prime} = -\frac{2}{3}W_{\alpha,n}(\phi) 
\end{equation}
solve the equations of motion \eqref{Eq-mov1}, \eqref{Eq-mov2} and \eqref{Eq-mov3}. Then, the potential in the curved spacetime reads
\begin{equation}
\label{Potential-Brane}
\mathcal{V}_{\alpha,n}(\phi) = \frac{1}{2} \left( \frac{dW_{\alpha,n}}{d\phi} \right)^2 - \frac{4}{3} W^2_{\alpha,n}(\phi).
\end{equation}

\begin{figure}[!t]
 \centering
    \includegraphics[width=0.45\textwidth]{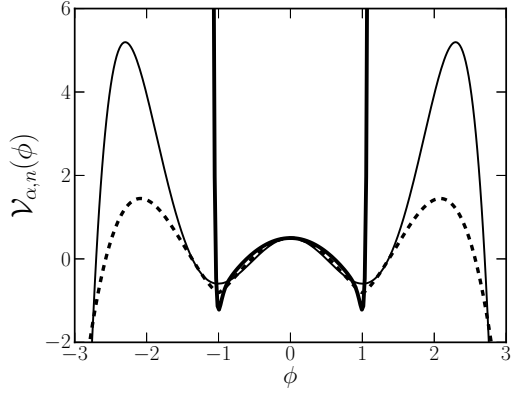}
 \caption{Potential in curved spacetime. The thin line corresponds to a brane being engendered by the $\phi^4$ model. The dashed line is the plot of $\mathcal{V}_\alpha$ and the thick line the plot of $\mathcal{V}_n$.}
 \label{Fig_Vbrana}
\end{figure}

In order to achieve $W_{\alpha,n}$, we performed the numerical quadrature of Eq. \eqref{Superpotential}. Then it was possible to construct the potentials that engender the brane. We plot in Fig. \ref{Fig_Vbrana} the potentials $\mathcal{V}_{\alpha}$ and $\mathcal{V}_n$. Note that only $\mathcal{V}_n$ possesses a compacted behaviour. The results for the model $\mathcal{L}_n$ agrees with the analytical solution for $W_n$ and $\mathcal{V}_n$ given, respectively, by
\begin{equation}\label{eq:supot}
W_n(\phi)=\phi - \frac{\phi^{2n+1}}{2n+1},
\end{equation}
and
\begin{equation}\label{eq:sgpot}
\mathcal{V}_n(\phi)= \frac{1}{8}(1-\phi^{2n})^2-\frac{1}{3} \left(\phi-\frac{\phi^{1+2n}}{1+2n}\right)^2.
\end{equation}

\begin{figure*}[!]
\subfigure[]{
\includegraphics[scale=0.3]{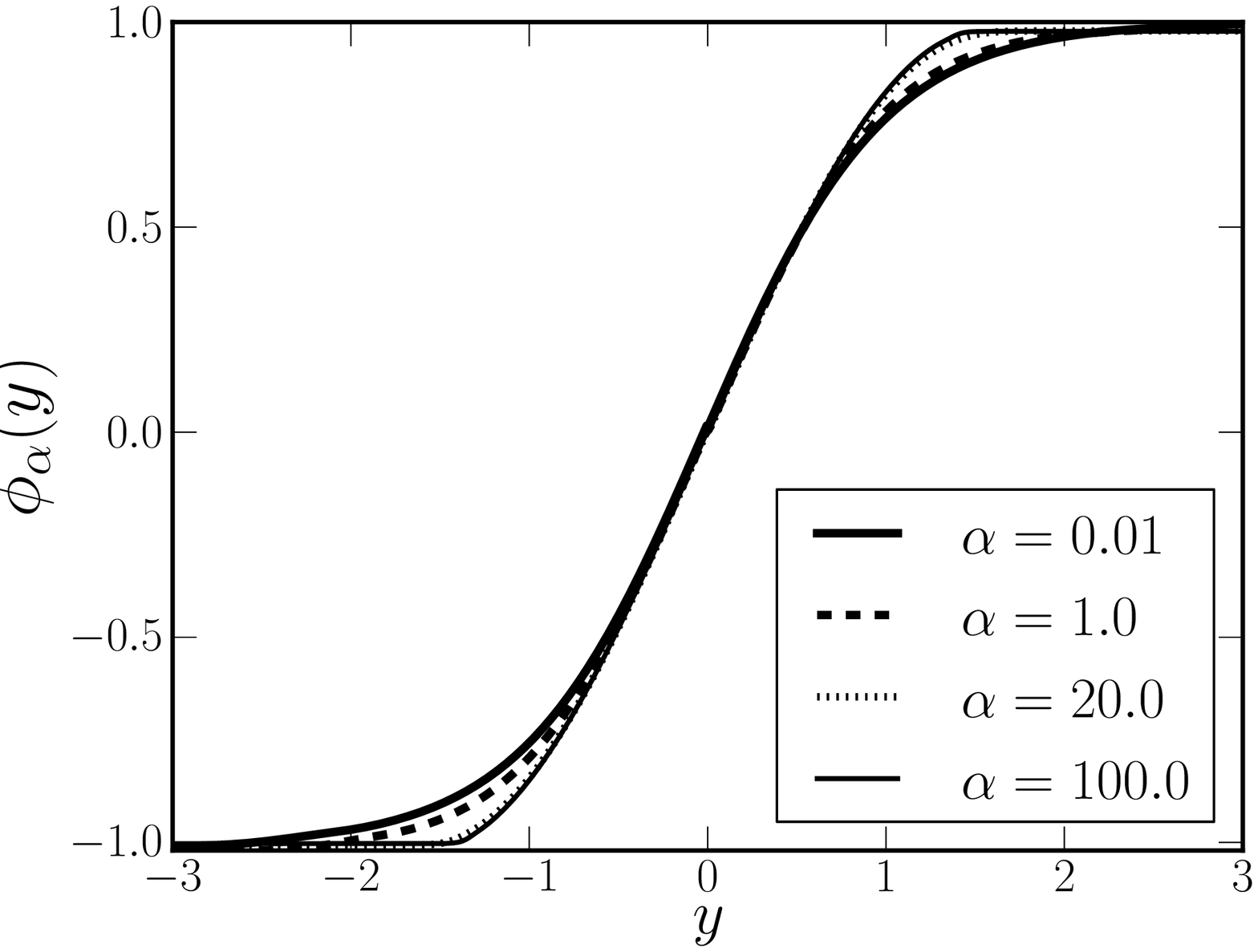}
\label{Fig_phi_alpha}}
\subfigure[]{
\includegraphics[scale=0.3]{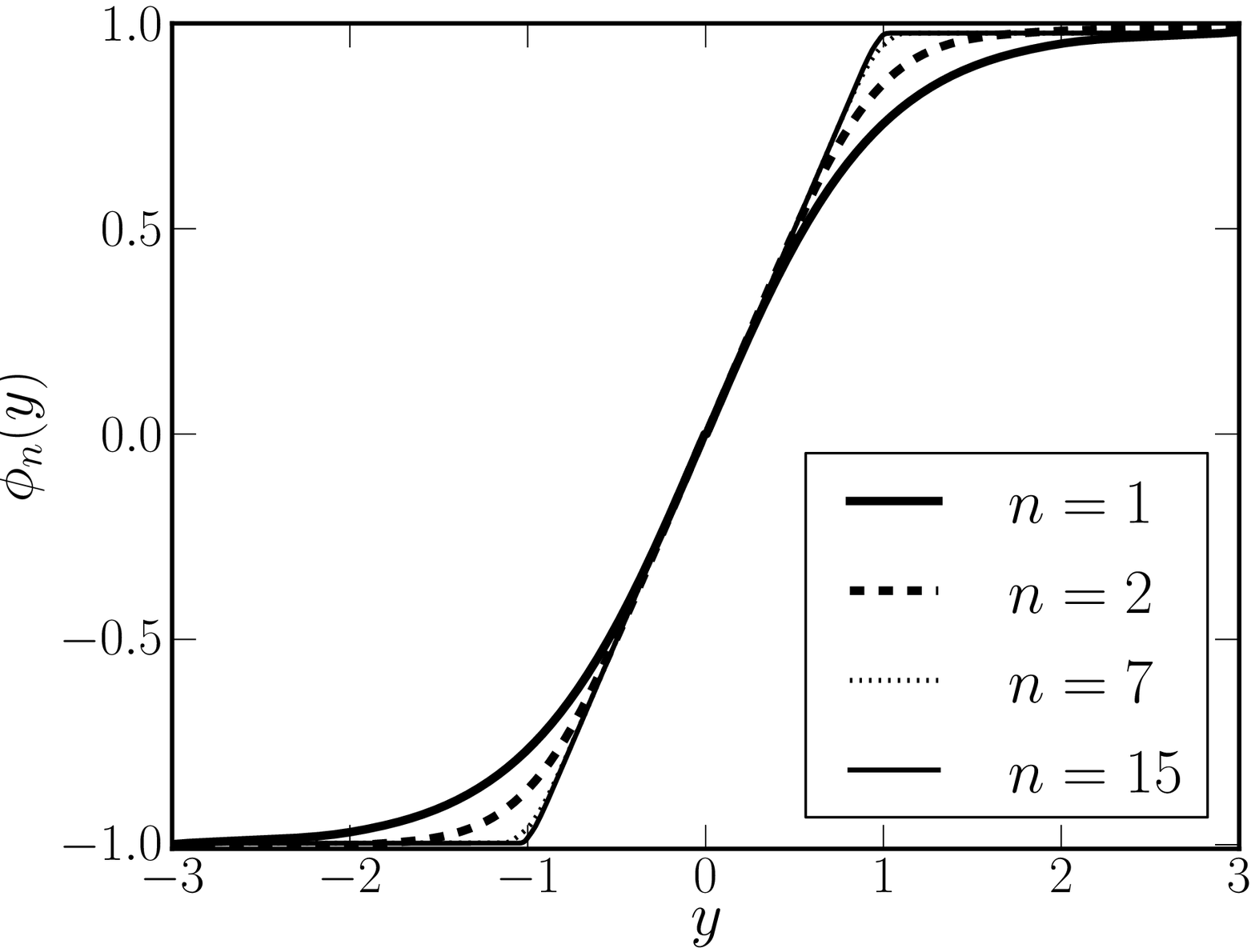}
\label{Fig_phi_n}}
\caption{Scalar field solution $\phi_\alpha(y)$ in (a) and $\phi_n(y)$ in (b) which engenders the brane. Note that the solution $\phi_n$ for large $n$ clearly resembles a compact defect.}
\label{Fig-phi}
\end{figure*}

\begin{figure*}[t]
\centering
\subfigure[]{
\includegraphics[scale=0.28]{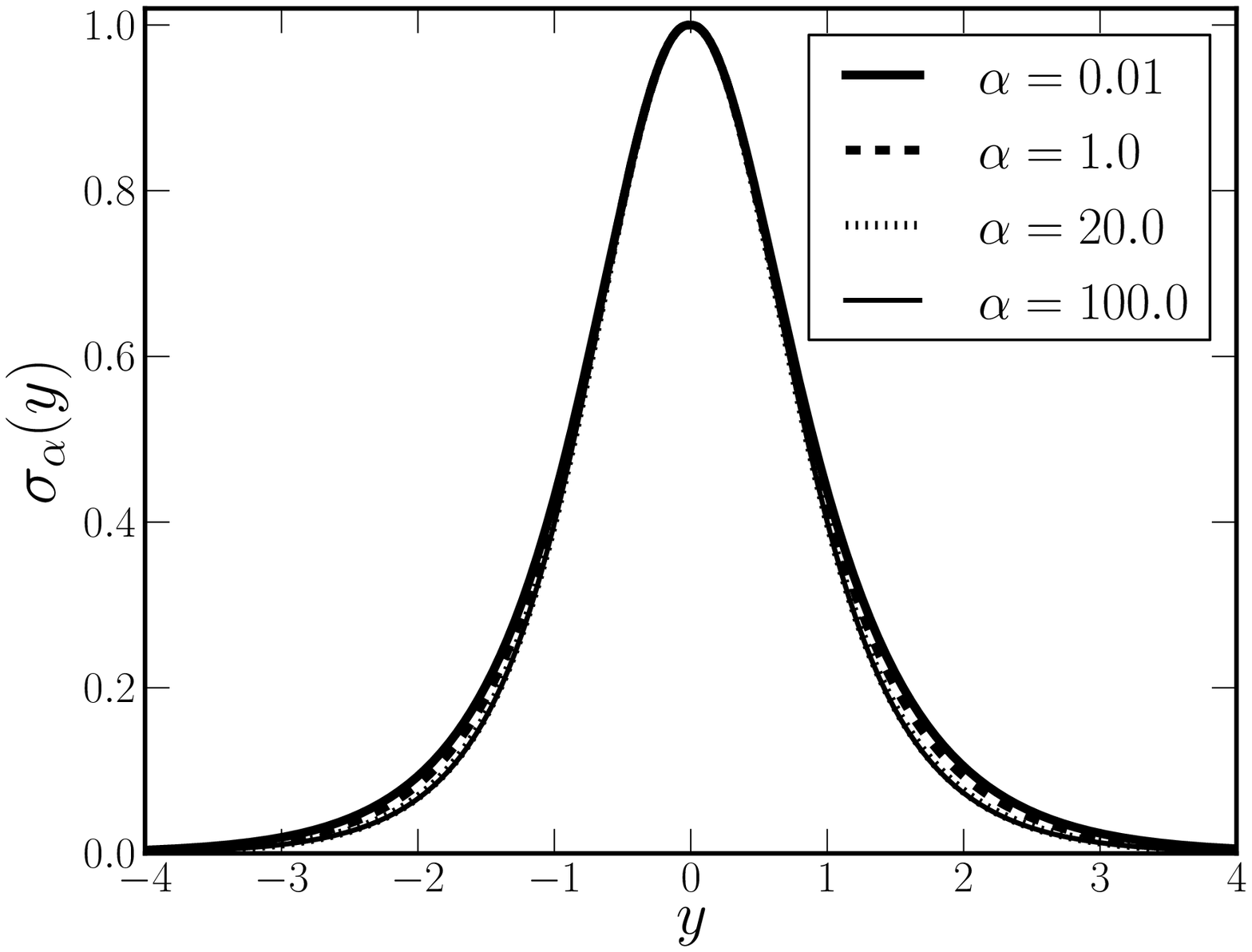}
\label{Fig_A_alpha}}
\subfigure[]{
\includegraphics[scale=0.28]{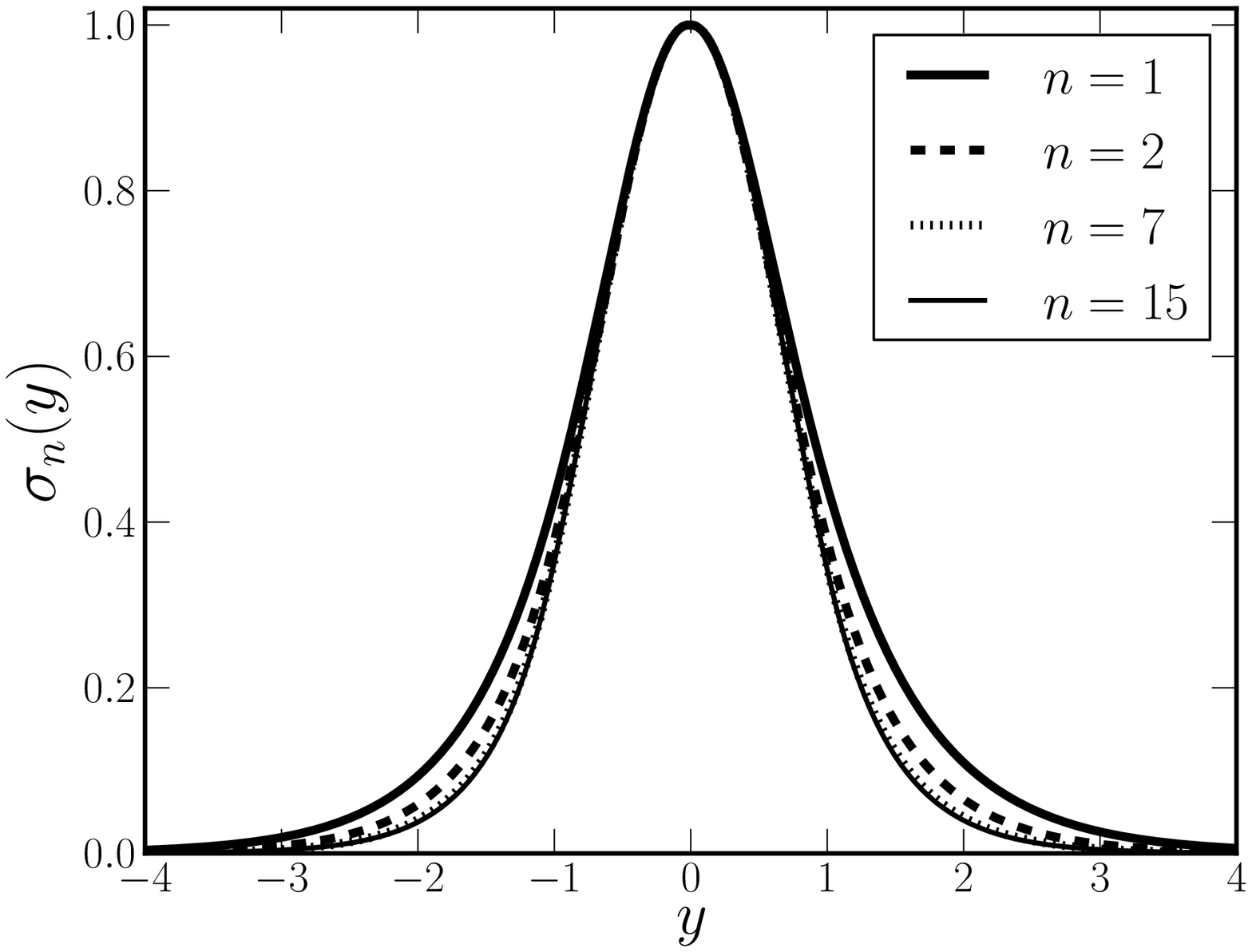}
\label{Fig_A_n}}
\caption{Warp factors $\sigma_{\alpha,n}(y) \equiv \e^{2A_{\alpha,n}(y)}$.}
\label{Fig_A}
\end{figure*}

With the numerical approximation of the superpotencial $W_{\alpha,n}(\phi)$ in hands, we were able to attain the scalar fields $\phi_{\alpha,n}(y)$ and the warp functions $A_{\alpha,n}(y)$. We solved the static equations of motion \eqref{FirstOrder_EqMotion} using fourth order Runge-Kutta algorithms imposing that $\phi_{\alpha,n}(0) = 0$. We plot in Fig. \ref{Fig-phi} the solutions $\phi_{\alpha,n}(y)$ for different values of the parameters. As for the potential functions, the solutions behave like kinks for $n = 1$ and $\alpha \rightarrow 0$. However, for very large values of the parameters, the solutions identify with the compacton soliton. Furthermore, note that $\phi_n(y)$ resembles much better to the compacton defect in the finite domain $[-1,1]$. To attain the warp functions, we employed the shooting method in Eq. \eqref{Eq-mov2} with Runge-Kutta routines. The warp factors $\sigma_{\alpha,n}(y) \equiv \e^{2A_{\alpha,n}(y)}$ are plotted in Fig. \ref{Fig_A}. Note that $\sigma_{\alpha}$ varies much slower than $\sigma_n$.

\begin{figure}[!h] 
       \begin{minipage}[t]{0.48 \linewidth}
           \includegraphics[width=\linewidth]{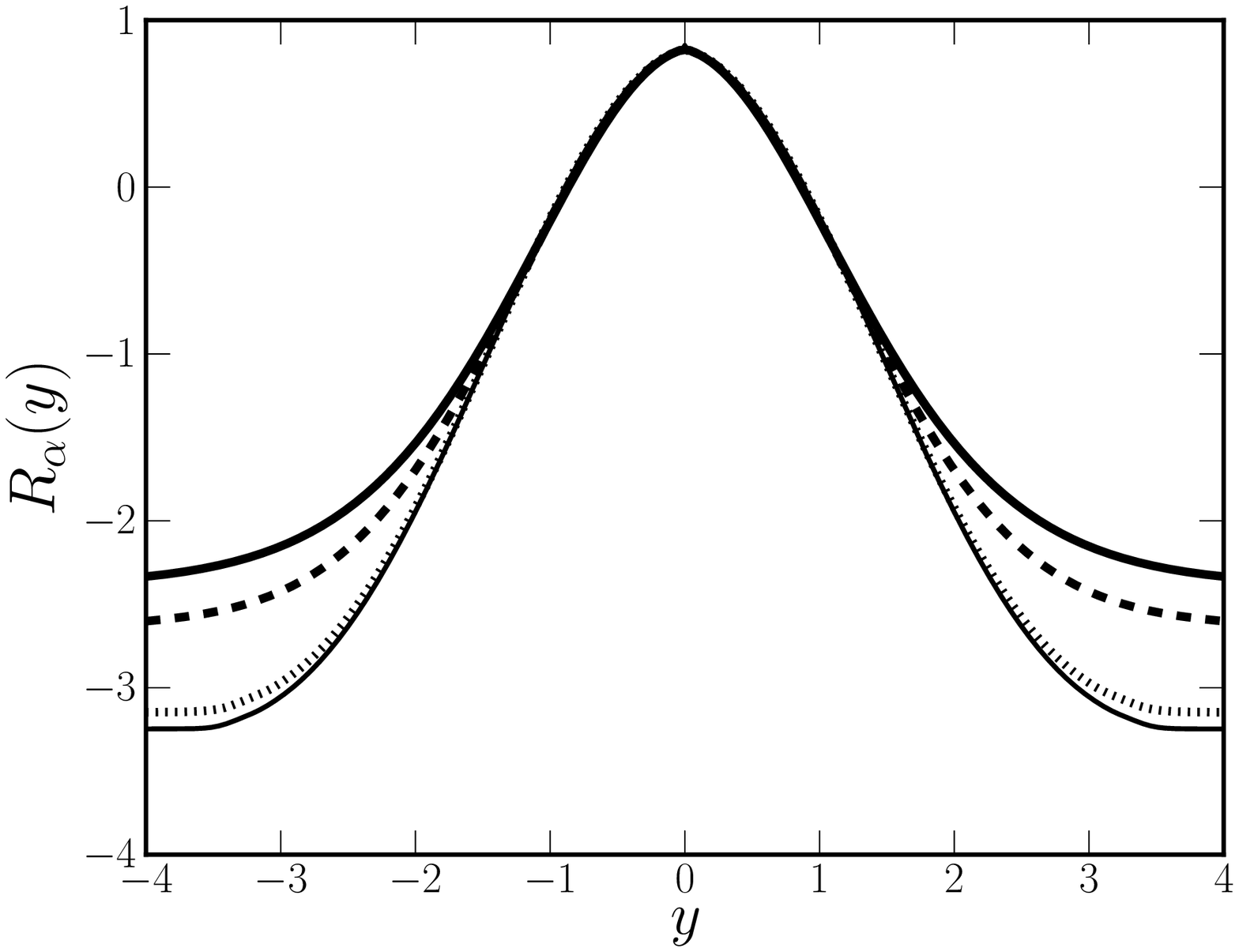}\\
           \caption{Scalar curvature $R_{\alpha}(y)$ for $\alpha = 0.01$ (thick line), $\alpha = 1.0$ (dashed line), $\alpha = 20.0$ (dotted line) and $\alpha = 150.0$ (thin line). }
          \label{Fig_R_alpha}
       \end{minipage}\hfill
       \begin{minipage}[t]{0.48 \linewidth}
           \includegraphics[width=\linewidth]{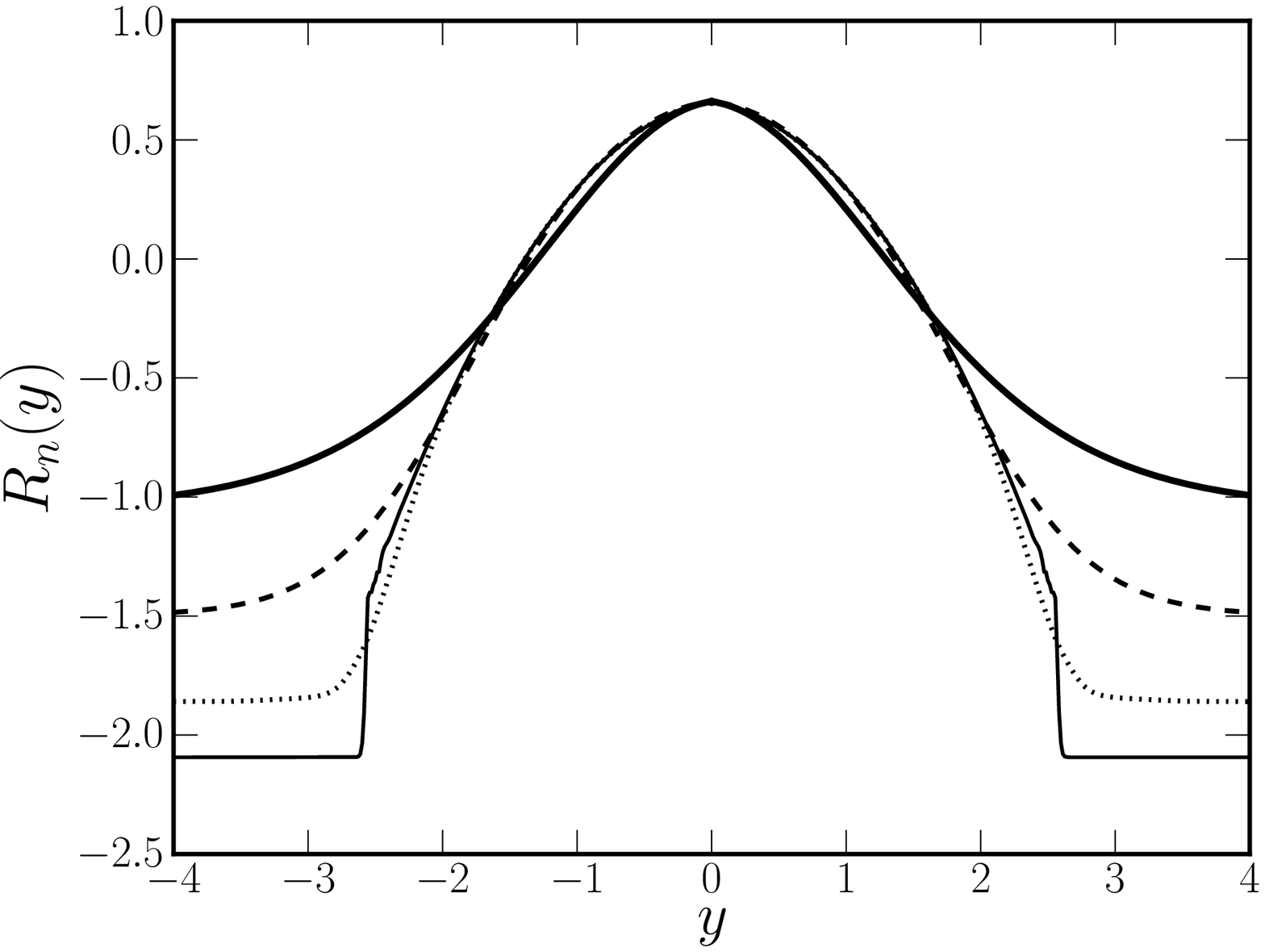}\\
           \caption{Scalar curvature $R_{n}(y)$ for $n = 1$ (thick line), $n = 2$ (dashed line), $n = 7$ (dotted line) and $n = 100$ (thin line). The sudden change to a negative constant value clearly characterizes the hybrid behaviour of the brane.}	
           \label{Fig_R_n}
       \end{minipage}
   \end{figure}

The hybrid behaviour of the brane is more clearly revealed in the scalar curvature, given by $R_{\alpha,n}(y) = -\left[ 8 A_{\alpha,n}^{\prime\prime}  + 20\left( A_{\alpha,n}^{\prime} \right)^2 \right]$. We plot in Figs. \ref{Fig_R_alpha} and \ref{Fig_R_n} the scalar curvatures $R_{\alpha}$ and $R_n$, respectively. In both cases, the $AdS_5$ limit for the bulk is present. However, note that $R_n$ has a sudden change to a constant negative value. Thus, the model $\mathcal{L}_n$ conducts to a \textit{hybrid} brane in a better way.

\section{Gravity fluctuations} 
\label{Sec-Localizacao}

We are interested in study the stability of the gravity sector of a braneworld scenario engendered by a compacton-like structure. Hence, we introduce the small perturbation $h_{\mu\nu}(\mathbf{x},y)$ as
\begin{equation}
\label{perturbation}
ds^2 = \sigma(y)(\eta_{\mu\nu} + h_{\mu\nu})dx^{\mu\nu}dx_{\mu\nu} + dy^2.
\end{equation}
Imposing the transverse-traceless gauge and $h_{5N} = 0$, the graviton equation of motion is \cite{Gremm}
\begin{equation}
\label{Graviton-Sturm}
h_{\mu\nu}^{\prime\prime} + \frac{ 2\sigma^{\prime}}{\sigma}h_{\mu\nu}^{\prime} = \sigma^{-1} \Box h_{\mu\nu},
\end{equation}
where $\Box$ is the $(3+1)-$dimensional d'Alembertian. Furthermore, assuming the Kaluza-Klein (KK) decomposition $h_{\mu\nu}(\mathbf{x},z) = \sum_k h_{\mu\nu}^{(0)}(\mathbf{x})\phi_k(z)$, where $\eta_{\mu\nu}\partial^{\mu}\partial^{\nu}h_{\mu\nu}^{(0)} = - m_k^2 h_{\mu\nu}^{(0)}$, with $m$ being the four-dimensional KK mass of the fluctuation, the gravitational KK modes in the extra dimension is described by the following Sturm-Liouville equation
\begin{equation}
\label{Graviton-Sturm-ExtraCoord}
\phi_k^{\prime\prime} + \frac{ 2\sigma^{\prime}}{\sigma}\phi_k^{\prime} = - m_k^2\sigma^{-1} \phi_k.
\end{equation}

To deal with a conformal metric, we change the coordinate to $dz = \sigma^{-1/2} dy$ \cite{Metastable}. Further, defining $\phi_k(y) =  \sigma^{-3/4}\psi_k(z)$,  the Sturm-Liouville equation \eqref{Graviton-Sturm-ExtraCoord} reduces to a Schr\"odinger-like form \cite{Metastable}
\begin{equation}
\label{Graviton-Schr}
-\ddot{\psi_k}(z) + U_{\alpha,n}(z)\psi_k(z) = m_k^2 \psi_k(z), 
\end{equation}
where $U_{\alpha,n}$ is the analogue quantum potential
\begin{equation}
\label{Potencial-Schr}
U_{\alpha,n}(z)= \frac{3}{4}\left[ \frac{\ddot{\sigma}_{\alpha,n}}{\sigma_{\alpha,n}} - \frac{1}{4} \left(\frac{\dot{\sigma}_{\alpha,n}}{\sigma_{\alpha,n}}\right)^2\right],
\end{equation}
and the over-dots represents derivatives with respect to $z$.
The Eq. \eqref{Graviton-Schr} does not concede tachyons as shown in Ref. \cite{Branalizacao}. Furthermore, the gravitational zero-mode ($m = 0$ normalizable state) is trapped in the brane and is given by $\psi_{m = 0}(z) = N \sigma^{3/2}(z)$, where $N$ is a  normalization constant\cite{Bazeia-JHEP}. 

\begin{figure}[!h] 
       \begin{minipage}[t]{0.48 \linewidth}
           \includegraphics[width=\linewidth]{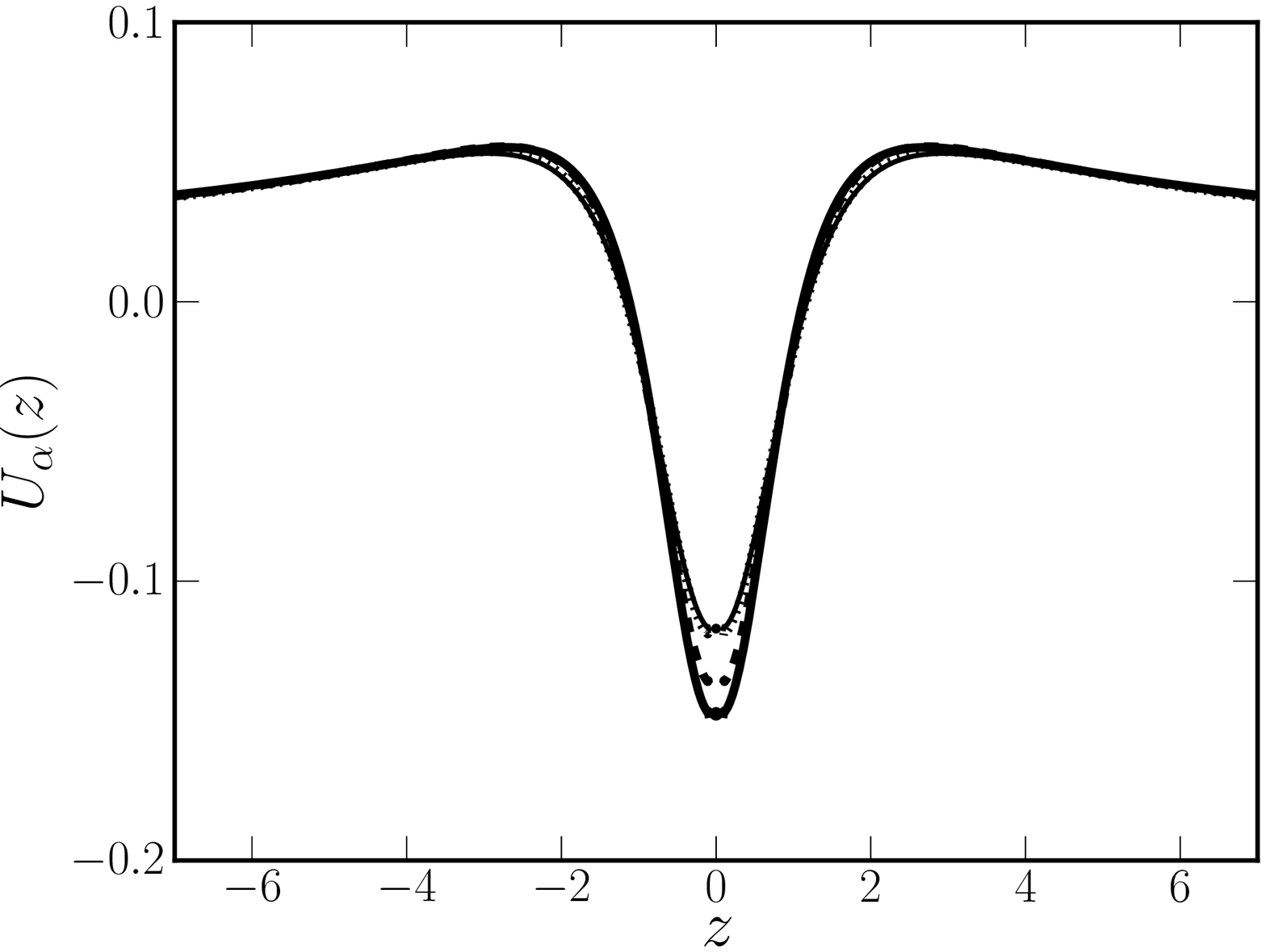}\\
           \caption{Schr\"odinger-like potential $U_{\alpha}(z)$ for $\alpha = 0.01$ (thick line), $\alpha = 1.0$ (dashed line), $\alpha = 20.0$ (dotted line), $\alpha = 150.0$ (thin line). There are no significant changes in the structure of the volcano-like potential with respect to $\alpha$. }
          \label{Fig_U_alpha}
       \end{minipage}\hfill
       \begin{minipage}[t]{0.48 \linewidth}
           \includegraphics[width=\linewidth]{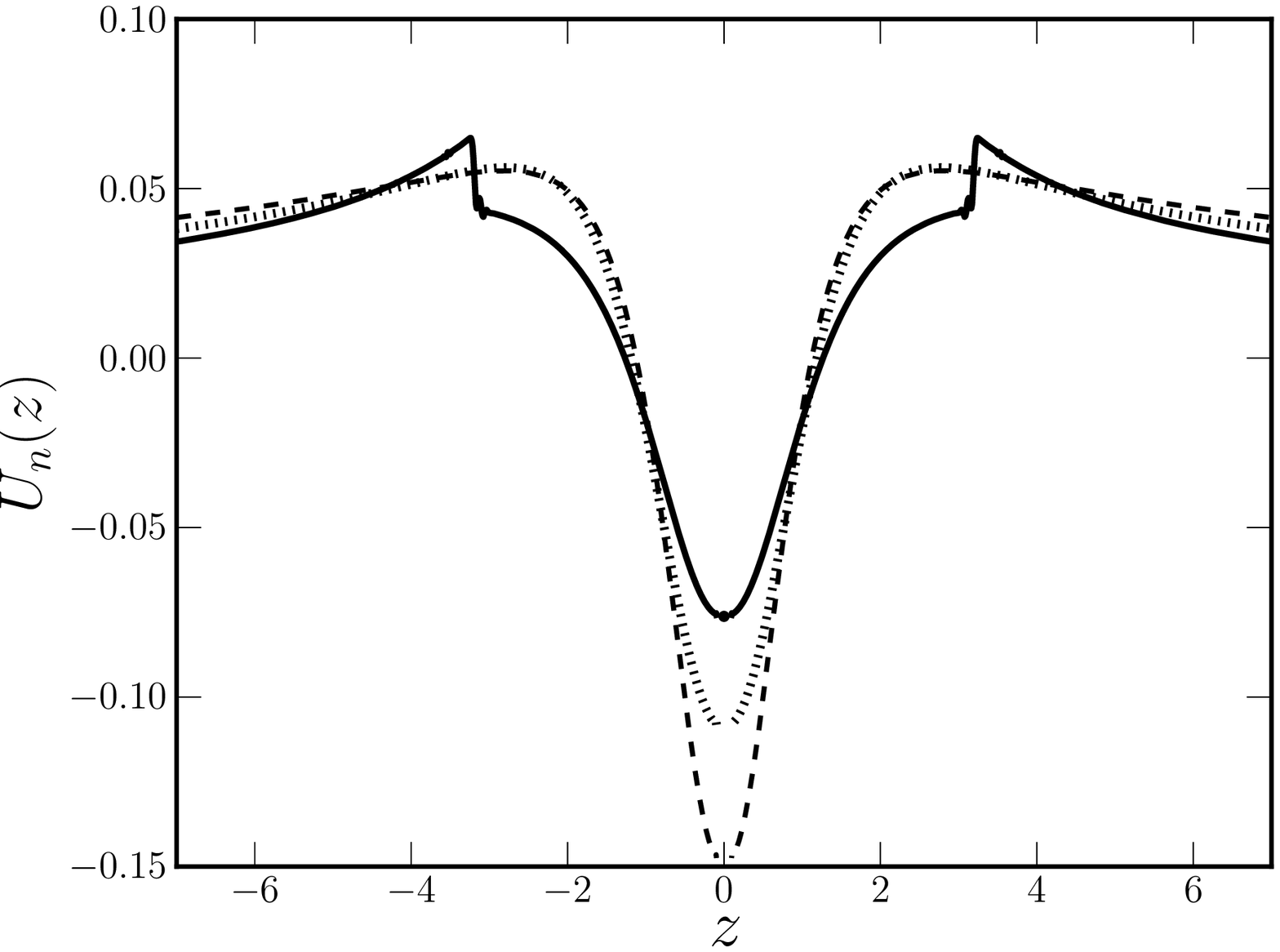}\\
           \caption{Schr\"odinger-like potential $U_{n}(z)$ for $n = 1$ (dashed line), $n = 2$ (dotted line) and $n = 100$ (thick line). The compact property of the scalar field solution causes the sudden change, evincing the hybrid brane profile.}	
           \label{Fig_U_n}
       \end{minipage}
   \end{figure}

We plot in Figs. \ref{Fig_U_alpha} and \ref{Fig_U_n} the potentials $U_{\alpha}$ and $U_{n}$, respectively. The potentials have the usual volcano shape \cite{Gremm}, which may support bound states. However, only the $U_n$ potential possesses the hybrid profile. Note that the thin brane pattern ($\sim z^{-2}$) \cite{RS2} is present forthwith after the compact domain ($\approx |z| \leq 3 $) instead of asymptotically. Hence, the model $\mathcal{L}_{\alpha}$ does not prompt significant changes in the usual thick brane scenario.


\section{Phenomenological Implications} 
\label{Sec-Phenomenologia}

An interesting and relevant point to analyze is the correction in the Newtonian potential due to the influence of the Kaluza-Klein modes. A solution of the Schr\"odinger-like equation, $\psi_{k}$ with energy $m_k^2$ contributes with a Yukawa-like correction to the Newton's law with a term $\frac{\psi_k^2(0)}{M_*^3}\frac{\e^{-m_kr}}{r}$, where $\psi_k(z)$ is normalized to $\int d z |\psi_{k}(z)|^{2} = 1$ and $M_{*}$ is the fundamental Planck scale in $5D$ \cite{Csaki-UniversalAspects}. If the analogue quantum potential goes to zero at infinity, there is a continuum of scattering states. However, for the case when $U(z) > 0$ for $|z| \rightarrow \infty$, the excited states are separated by a gap from the massless mode \cite{Csaki-UniversalAspects}. Hence, the gravitational potential between two point-like sources of mass $M_1$ and $M_2$ located at the origin ($z = 0$) in the transverse space, will be exponentially suppressed as
\begin{equation}
\mathcal{U} (r) \simeq G\frac{M_1 M_2}{r} + M_{*}^{3}M_1 M_2 \sum_k\frac{\e^{-m_kr}}{r}\psi_{k}^{2}(0).
\label{NewtonCorrection}
\end{equation} 

In order to study quantitatively the effects of this correction, it is necessary to attain the mass spectrum $\{m_k\}$. To this aim, a numerical procedure is needed to solve the equations of the massive modes. Fortunately, the matrix method \cite{MatrixMethod} is well applicable to Sturm-Liouville problems since it well approximates the first eigenvalues, which are of interest. This technique was successfully applied in codimension-2 models for the gravitational \cite{DiegoGraviton}, gauge \cite{DiegoGauge} and fermionic  fields \cite{DiegoFermion}. We discretized the Eq. \eqref{Graviton-Sturm-ExtraCoord} and the boundary condition $\phi^{\prime}(-\infty) = \phi^{\prime}(\infty) = 0$ in the domain $[-10.0, 10.0]$ with uniform step-size $h = 0.01$ up to second order truncation error. We performed the numerical analysis for the cases $n = 1$ (thick domain wall brane) and $n = 100$ (hybrid brane), since the model $\mathcal{L}_\alpha$ does not differ significantly from the domain wall case. In both cases, the spectrum exhibited the linearly increasing pattern as usual from the Kaluza-Klein theories.  

Using the Numerov algorithm \cite{Numerov}, we solved Eq. \eqref{Graviton-Schr} for all mass eigenvalues obtained previously by the Matrix method. The Numerov method is well-known of gravity localization on thick branes \cite{AdaltoNumerov}. In order to set free the values of the wave-function in the origin, we relinquished the \textit{unphysical} boundary condition \cite{Gremm} adopting the conditions $\psi^{\prime}(-\infty) = \psi^{\prime}(\infty) = 0$. This condition allows different contributions of the massive modes to the correction in the Newton's law favouring the resonant states.

It is important to mention that resonant states can happen for some particular energies, where incident plane waves can resonante with the potential $U(z)$ and consequently have a disproportionately large value of $\psi_k(0)$ \cite{CASA-Ressonancia}.  However, the deformation parameter $n$ does not regulate $U(z)$ considerably. Note that, by construction, the hybrid brane does not possesses a parameter related to the thickness of the brane which is responsible to control the height of the barrier and the width of the potential well. This feature makes difficult the search for resonant states. We have not found resonant states using the well-succeeded resonance methods \cite{CASA-Ressonancia, Chineses-Ressonancia}.

\begin{figure}[h]
 \centering
    \includegraphics[width=0.45\textwidth]{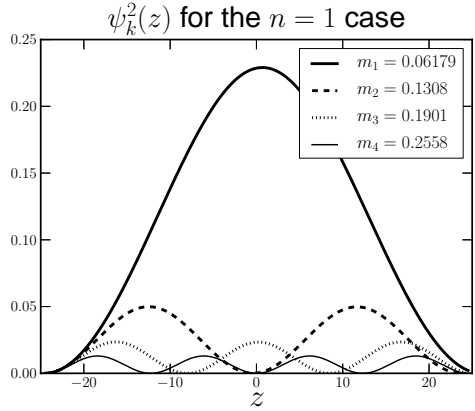}
 \caption{Normalized numerical solutions of the Schr\"odinger-like equation for the first four eigenvalues for $n=1$ . Although no resonant states were found, the first eigenstate will contribute highly for the correction to the Newton's law. Moreover, some eigenfunction will give trivial contribution, since $\psi^2(0) = 0$. The solutions for higher values of $n$ did not presented considerable changes.}
 \label{Fig_FuncaoDeOnda}
\end{figure}

We present the numerical solutions of the Schr\"odinger-like equation for the first four eigenvalues in Fig. \ref{Fig_FuncaoDeOnda}. A noteworthy result is that the first eigenstate will contribute highly for the correction to the Newton's law. The subsequent normalized solutions present diminishing amplitudes. Moreover, some eigenfunctions (the $k$-odd solutions) will give trivial contribution, since $\psi^2(0) = 0$. The spectrum and wave functions did not present notable changes for variations of the parameter $n$. This is in accordance with the fact that bound states can happen for masses $m^2$ up to the maximum value of the potential barrier \cite{Csaki-UniversalAspects,Csaki-Quasilocalization}. Note that the maxima of $U_n(z)$ in Fig. \ref{Fig_U_n} are very close for $n=1$ and $n=100$. Hence we only present graphs for $n = 1$.

\begin{figure}[h]
 \centering
    \includegraphics[width=0.45\textwidth]{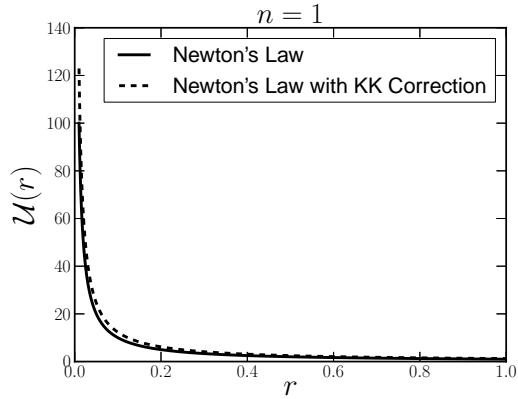}
 \caption{Newtonian potential  with the correction due to Kaluza-Klein modes for $n = 1$ for two unitary masses.  It can be seen that the gravitational force is slightly increased at short distances due to the massive modes.}
 \label{Fig_Newton-Correction}
\end{figure}

With such results, we are able to evaluate the correction to the Newton's law given by Eq. \eqref{NewtonCorrection}. To this, we simplified the Planck mass to $G = M_{*} = 1$. We plot in Fig. \ref{Fig_Newton-Correction} the slight deviation of the gravitational law due to the Kaluza-Klein modes. We may conclude that the gravitational force is slightly increased at short distances. Moreover, the phenomenology is not different if the braneworld is engendered by a kink or a compacton defect.

\section{Conclusions \label{Sec_Conclusion}}

We studied the gravity localization on codimension-1 brane worlds engendered by compact-like kinks, the so-called \textit{hybrid brane}. In such scenario, the thin brane behaviour is present when the extra dimension is outside the compact domain where the energy density is non-trivial instead of asymptotically as in the usual thick brane models. In the literature, a mechanism which transforms smoothly kinks to compactons was proposed in Ref. \cite{Bazeia-FromKinks} using two specific models, $\mathcal{L}_{\alpha}$ and $\mathcal{L}_n$. In such reference, the hybrid brane was constructed using only the second model. In this paper, we used suitable numerical methods to construct branes from a compact-like defect using both models and studied their gravity localization. We showed that the model $\mathcal{L}_n$ conducts to the hybrid brane scenario in a clearer way. Firstly, the scalar field in the curved spacetime $\phi_n(y)$ engenders the compact-like behavior much clearer. Moreover, the scalar curvature $R_n$ has a \textit{sudden} change to a constant negative value (which characterizes the $AdS_5$ limit for the bulk) clearly revealing the hybrid behavior of the brane. 

The study of the gravity fluctuations showed that the zero-mode is trapped in the brane, as desired. Furthermore, the analogue quantum potential revealed that the model $\mathcal{L}_{\alpha}$ does not present significant changes from the usual thick brane derived from a kink defect. Since the hybrid brane does not possess a parameter related to its thickness, which is responsible to control the height of the barrier and the width of the potential well, we have not found resonant states using resonance methods.

The noteworthy results of the paper lay in the phenomenological implications of the massive modes. We studied quantitatively the corrections to the gravitational potential between two point-like sources of mass. By means of suitable numerical methods, we attained the mass spectrum of the graviton. The usual linearly increasing behaviour from the Kaluza-Klein theories were recovered for small masses which are of interest. We therefore, used such discrete mass spectrum to solve the correspondent Schr\"odinger equation. We showed that the first eigenstate contributes highly for the correction to the Newton's law. The subsequent normalized solutions present diminishing amplitudes. Moreover, odd eigenfunctions will give trivial contribution, since the value of the wave function at the origin in the transverse space, where the core of the brane is located, is null. With the above results, we were able to evaluate the correction to the Newton's law due to the Kaluza-Klein tower. We concluded that the gravitational force is slightly increased at short distances, and the phenomenology is not different if the braneworld is engendered by a kink or a compact defect. Such results may be used to match phenomenological measurements of the gravitational law in particle colliders to probe warped extra dimensions.

The behavior of the scalar curvature that we depicted in Fig.~7 encourages us to go further on this issue, and investigate other braneworld scenarios, in particular the case where the brane itself tends to become compact along the extra dimension, as suggested in the very recent work of Ref. \cite{BM}.

\section{Acknowledgments}

The authors thank D. Bazeia for useful discussions and thank the Coordena\c{c}\~{a}o de Aperfei\c{c}oamento de Pessoal de N\'{i}vel Superior (CAPES), the Conselho Nacional de Desenvolvimento Cient\'{i}fico e Tecnol\'{o}gico (CNPQ), and Funda\c{c}\~{a}o Cearense de apoio ao Desenvolvimento Cient\'{i}fico e Tecnol\'{o}gico (FUNCAP) for financial support.

\end{document}